\documentclass[final,3p,times]{elsarticle} 

\usepackage{amssymb}

\usepackage{setspace}

\usepackage{hyperref}
\usepackage{graphicx}
\graphicspath{{figs/}}
\usepackage{color}
\usepackage{epstopdf}
\usepackage{tabularx}
\usepackage{subfig}
\usepackage{longtable}
\usepackage{gensymb}
\usepackage[font=small,labelfont=bf]{caption}

\usepackage{amssymb}
\usepackage{amsmath}
\usepackage{multirow}
\usepackage{array}
\journal{Journal name}

\begin{document}
	\doublespacing

\begin{frontmatter}



\vspace*{2mm}

\title{{Architecting materials for extremal stiffness, yield and buckling strength}}


\author[]{Fengwen Wang\corref{cor1}}
\ead{fwan@dtu.dk}

\author[]{Ole Sigmund}

\cortext[cor1]{Corresponding author}

\address{Department of Civil and Mechanical Engineering, Technical University of Denmark, Koppels Allé, Building 404, 2800 Kgs. Lyngby, Denmark}


\begin{abstract}
{This paper proposes a methodology for architecting microstructures with extremal stiffness, yield, and buckling strength  using topology optimization. The optimized microstructures reveal an interesting transition from simple lattice like structures for yield-dominated situations to hierarchical lattice structures for buckling-dominated situations. The transition from simple to hierarchical is governed by the relative yield strength of the constituent base material as well as the volume fraction. The overall performances of the optimized microstructures indicate that  maximum  strength is determined by the buckling strength at low volume fractions and yield strength at higher volume fractions, regardless of the base material's relative yield strength. {The non-normalized properties of the optimized microstructures show that  higher base material  Young's modulus leads to both higher Young's modulus and strength of the architected microstructures. Furthermore, the polynomial order  of the maximum strength lines with respect to mass density obtained from the optimized microstructures reduces as base material relative yield strength decreases,  reducing from 2.3 for buckling dominated Thermoplastic Polyurethane  to 1 for yield dominated steel microstructures.}}

\end{abstract}


\end{frontmatter}





\section{Introduction} \label{sec1}
Exploring novel material architectures with extremal properties has been a constant quest in the field of  material design and lightweight engineering. These developments have been further promoted by advances in additive manufacturing facilitating  fabrication of functional materials with unprecedented complexity~\citep{Meza2015, Zheng2016}. As substitutions to trial and error and limited human intuition, topology optimization methods have been shown to be powerful tools in designing novel materials in various applications~\citep{Sig00,Bendsoe2003,OsaGue16}. Up to now, however, the field of architected materials has focused on individual material properties like stiffness, yield or buckling strength but less so on the intricate trade-off between these properties.


Material stiffness and strength are fundamental properties for determining material load-bearing capacity as they measure  the material’s deformation resistance and ultimate load-carrying capacity, respectively. The quest for optimal stiff and strong materials depends on many aspects, including the loading conditions and constituent base materials. Stiffness-optimal materials {meeting} theoretical upper bounds have been obtained via systematic design approaches~\citep{Sigmund1995, Guest2006, Huang2011, Andreassen2014, Berger2017}.  It has been shown that plate microstructures reach the Hashin-Shtrikman bounds~\citep{Hashin1962} in the low volume fraction limit and remain within 10\% of the theoretical upper bounds at moderate volume fractions.  A few studies have focused on improving microstructure strength, including stress minimization and buckling strength optimization. It has been shown that the maximum von Mises {stress} can be  reduced when taking stress into account during the microstructure optimization procedure either as constraints or objectives~\citep{Ferrer2021,Collet2018,Coelho2019,Alacoque2021}. The stress singularity issue, where the stress at a point approaches infinity as the density at that point approaches zero, was handled by stress relaxation methods~\citep{Cheng1997,Duysinx1998,Bruggi2008,Le2010}. Large numbers  of local stress constraints can be aggregated to a global quantity using the p-norm, Kresselmeier–Steinhauser (KS)~\citep{Kreisselmeier1980} methods or {by the augmented Lagrangian method~\citep{SilAagSig21}}. Microstructure buckling strength optimization using topology optimization was first studied by~\cite{Neves2002}, focusing on cell-periodic buckling modes. More recently, 2D and 3D material microstructures were systematically designed to enhance buckling strength based on linear material analysis. This approach evaluates  effective material properties using the homogenization method~\citep{Hassani1998a} and material buckling strength under a given macro stress using linear buckling analysis (LBA) with Bloch-Floquet boundary conditions to capture all the possible buckling modes~\citep{Triantafyllidis1993,Thomsen2018,Wang2021}. Both studies showed that the optimized materials possess several times higher buckling strength than their references at the cost of {some stiffness degradation}. Subsequent 2D experimental verifications have further verified the buckling superiority of the optimized microstructures and validated the linear material evaluation~\citep{Bluhm2021}.

So far, optimized microstructures have been designed considering material buckling or yield failure separately. However, yield-optimized microstructures tend to be vulnerable to buckling failure. On the other hand, buckling-optimized microstructures assume constituent base materials with relatively high yield strength (yield strength to Young’s modulus ratio, $\sigma_1/E_1$), e.g., elastomers. For other base materials with low relative yield strengths, e.g., steel, the buckling-optimized microstructures fail due to localized yield. Hence the optimized strength superiority in both cases may significantly degrade in real applications. Furthermore, a previous numerical study showed that for a given microstructure topology, the failure mechanism switches from buckling- to yield-dominated failure as the volume fraction increases~\citep{Andersen2021}. Hence, it is crucial to consider both failure mechanisms  in the design procedure  considering different volume fractions. Furthermore, it is essential to provide microstructure candidates with programmable properties working for various base materials  and volume fractions fitting various applications.

In this study, we extend the stiffness/buckling studies from~\cite{Thomsen2018} and~\cite{Wang2021} to also include yield strength. Yield strength and  Young’s modulus share a monotonic relation in the optimized designs, as shown in the result section. Hence microstructure will be designed systematically by maximizing buckling strength with different yield strength bounds and considering different volume fractions. The optimized microstructure are further evaluated for different practical base materials ranging from low relative yield strength steel to high relative yield strength thermoplastic polyurethane (TPU). 

The paper is organized as follows. Section 2 summarizes basic formulations to evaluate microstructure stiffness and strength,  and formulates the optimization problem for designing 2D microstructures with extremal properties. Section 3 validates the proposed approach and presents topology-optimized microstructure sets for different volume fractions and corresponding performances considering different base materials. Finally, section 4 concludes the study.

\section{Optimization problem for 2D architected materials with enhanced stiffness and  strength} 
This section summarizes the basic formulations for designing 2D {architected materials}  with enhanced stiffness and strength using topology optimization. The finite element method is combined with homogenization theory and LBA to evaluate material  properties~\citep{Cook2002}. To accurately represent the stress situation, we employ the incompatible elements from~\cite{Wilson1973} and~\cite{Wilson1990}, i.e.,   the so-called $Q_{6}$ element.    Two additional so-called incompatible modes are considered to represent bending deformations accurately. The reader is referred to the work by~\cite{Wilson1973} and~\cite{Wilson1990} for additional formulations for the $Q_{6}$ elements.

\subsection{Material stiffness and strength evaluations}
\begin{figure}[!htb] 
		\includegraphics[width=1\textwidth]{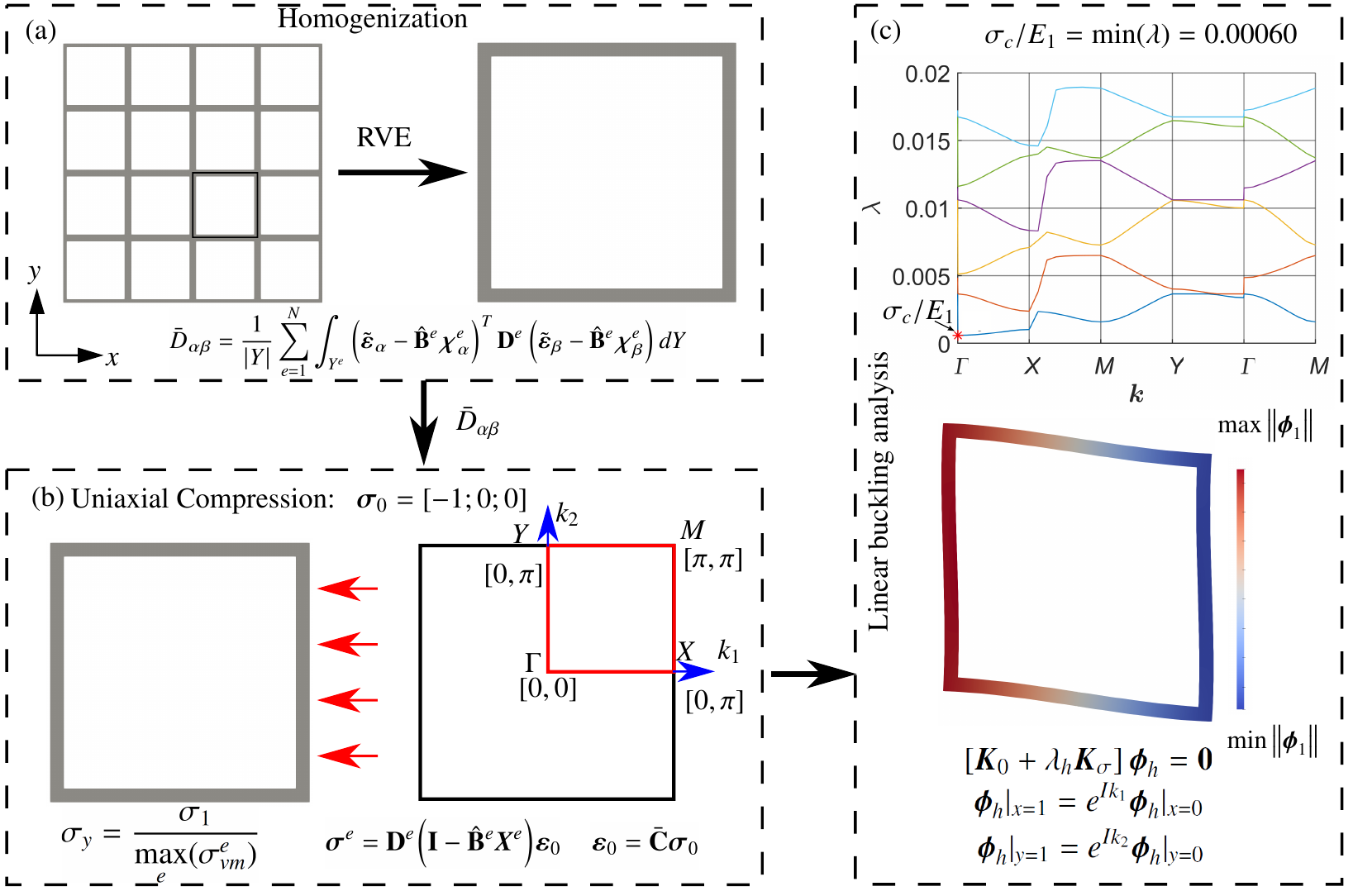} 
	{\caption{Flowchart for material stiffness and strength evaluations based a stiffness-optimal orthotropic material with a volume fraction of 0.2.  (a) Homogenization of a periodic material  using a representative volume element (RVE), i.e., microstructure. (b) Calculation of yield strength and illustration of the corresponding  irreducible  Brillouin zone (IBZ)  using uniaxial compression. (c) Bucking band structure, buckling strength and mode.     }	\label{fig:Illustration}}
\end{figure}

Under small strain assumptions, the material stiffness and strength are evaluated using the homogenization
approach and LBA together with Bloch-Floquet theory to account for buckling modes
at different wavelengths based on the periodic microstructure.  Fig.~\ref{fig:Illustration} summarizes the  material stiffness and strength evaluation procedure.  

The microstructure is assumed of unit size here. The symmetry properties of the effective elasticity matrix are exploited to represent the equations in a more compact form using  the abbreviation $kl\rightarrow \alpha$: $11\rightarrow 1$, $22\rightarrow 2$,  $(12,21)\rightarrow 3$.  The  effective elasticity matrix is calculated by an equivalent energy-based homogenization formulation~\citep{Sigmund1995,Hassani1998a} via
\begin{align}\label{eq:chi_disc}   
	\bar{D}_{ \alpha\beta} = \frac{1}{|Y|} \sum_{e=1}^N   \int_{Y^e} \left(\tilde{\boldsymbol{\varepsilon}}_{\alpha} - 	\hat{\mathbf{B}}^e\boldsymbol{\chi}^e_{\alpha} \right)^T \mathbf{D}^e \left(\tilde{\boldsymbol{\varepsilon}}_{\beta} - \hat{\mathbf{B}}^e\boldsymbol{\chi}^e_{\beta} \right) dY,\nonumber   \\
	\mathbf{K}_0 \boldsymbol{\chi}_{\alpha} = \mathbf{f}_{\alpha}, \quad \alpha=1,2,3,\\
	\boldsymbol{\chi}_{\alpha}|_{x=1}=\boldsymbol{\chi}_{\alpha}|_{x=0},   \quad   \boldsymbol{\chi}_{\alpha}|_{y=1}=\boldsymbol{\chi}_{\alpha}|_{y=0}.\nonumber  
\end{align}
Here $|Y|$ denotes the volume of the microstructure, $\sum$ represents a finite element assembly operation over all $N$ elements,  the superscript $ \left( \right) ^T$ denotes the transpose,  {$\hat{\mathbf{B}}^e$} with a size of $3\times 8$  is the condensed strain-displacement matrix of element $e$ in the $Q_{6}$ element formulation,  $\mathbf{D}^e$ is the elasticity matrix of the material in element  $e$,  $\tilde{\boldsymbol{\varepsilon}}_{\alpha}=\left[ \delta_{\alpha\beta}\right] $ denotes the 3 independent unit strain fields,   $\mathbf{f}_{\alpha}$ is the condensed equivalent  load vector induced by the $\alpha$th unit strain and $\mathbf{K}_0$  is the global  condensed elastic stiffness matrix.   The detailed formulation of $ \mathbf{K}_0$  and $\mathbf{f}_{\alpha}$  can be found  in~\cite{Wilson1990} and~\cite{Thomsen2018}. 

The effective material compliance matrix is $\bar{\mathbf{C}}=\bar{\mathbf{D}}^{-1} $.  {Under plane stress} assumptions, the effective Young's and bulk moduli are  calculated using  the effective elasticity or compliance matrices, stated as 
\begin{align}\label{eq:prop}
	\bar{E}=\frac{1}{\bar{C}_{11}},  \quad   
	\bar{\kappa} =\frac{  \bar{D}_{11}+\bar{D}_{12}}{2}.  
\end{align} 

For a prescribed  macroscopic stress state, $\boldsymbol{\sigma}_0$, the corresponding stress state of element $e$  is obtained by the superposition of the three perturbation fields induced by the  three unit strain fields in Eq.~\eqref{eq:chi_disc}, expressed as   
\begin{align}\label{eq:strain}
	\boldsymbol{\sigma}^e =
	\mathbf{D}^e \Big( \mathbf{I} - \hat{\mathbf{B}}^e 
	\boldsymbol{X}^e
	\Big)   \boldsymbol{\varepsilon}_0,  \  \rm{with} \  \boldsymbol{\varepsilon}_0 =   \bar{\mathbf{C}} \boldsymbol{\sigma}_0,
\end{align} 
where $\boldsymbol{\sigma}_0=  \boldsymbol{n}$ with $\boldsymbol{n}$ indicating the loading direction,  $\boldsymbol{n}=[-1,0,0]^T$ for the uniaxial loading case,  $\boldsymbol{X}^e= \left[\boldsymbol{\chi}^e_{1}\ \ \boldsymbol{\chi}^e_{2}\ \   \boldsymbol{\chi}^e_{3}\right] $ is a $8\times 3$ matrix containing the element perturbation fields.

The yield strength of the microstructure, ${\sigma}_y$,  is determined by the ratio between the base material yield strength ($\sigma_1$) and  the maximal von Mises stress in the microstructure, given as 
\begin{align}
	{\sigma}_y =\frac{{\sigma}_1 }{\max\limits_e(\sigma^e_{vm})}.
	\label{eq:Yield}
\end{align}
Here $\sigma^e_{vm}$ is the elemental von Mises stress, calculated  by
\begin{align}
	\sigma^e_{vm}=\sqrt{  \sigma^e_1\sigma^e_1 -\sigma^e_1\sigma^e_2+\sigma^e_2\sigma^e_2 +3\sigma^e_3\sigma^e_3  } =\sqrt{ \left( \boldsymbol{\sigma}^e\right)^T  \boldsymbol{M}\boldsymbol{\sigma}^e}, \label{eq:sigvm}
\end{align}
where $\boldsymbol{M}= \left[ \begin{array} {l l l}
	1 & -1/2 & 0 \\
	-1/2 &  1 & 0 \\ 
	0&  0 &3\end{array} \right] $

Based on the stress distribution in the microstructure,   subsequent LBA is performed  to evaluate the material buckling strength. Both short  and long wavelength buckling is captured by employing the Floquet-Bloch boundary conditions in the LBA~\citep{Neves2002,Thomsen2018,Triantafyllidis1993,}, stated as
\begin{align}\label{eq:buckling}
	\left[  \boldsymbol{K}_0 +   \lambda_h  \boldsymbol{K}_{\sigma}  \right]\boldsymbol{\phi}_h = \boldsymbol{0}, \\
	\boldsymbol{\phi}_h|_{x=1}=e^{Ik_1}\boldsymbol{\phi}_h|_{x=0},\quad  \boldsymbol{\phi}_h|_{y=1}=e^{Ik_2}\boldsymbol{\phi}_h|_{y=0}. \nonumber
\end{align}	
Here  $\boldsymbol{K}_{\sigma}=\sum_e \boldsymbol{K}^e_{\sigma} $  is the stress stiffness matrix with $\boldsymbol{K}^e_{\sigma} $ being the elemental stress stiffness matrix, $I=\sqrt{-1}$ is the imaginary unit, the smallest eigenvalue, $\lambda_1$, is the critical buckling strength for the given wave vector  ($\boldsymbol{k}=[k_1,k_2]^T$),  and $\boldsymbol{\phi}_1$ is the associated eigenvector. 

The material buckling strength, $\sigma_{c}$, is determined by the smallest eigenvalue for all the possible wave vectors in Eq.~\eqref{eq:buckling}, located in the first  Brillouin zone, $\lambda_{min}$,   i.e., $\sigma_{c}=\min\limits_{h,\boldsymbol{k}} \lambda_h$.    The associated eigenvector is the critical buckling mode.  The first  Brillouin zone is  the primitive cell in reciprocal space~\citep{Brillouin1953}, spanning over  $k_j \in [-\pi,\ \pi], \ j=1,2$. Previous studies have shown that the critical buckling mode can be captured by sweeping $k$-vectors along the boundaries of the  irreducible  Brillouin zone (IBZ)~\citep{Thomsen2018,Geymonat1993} that is determined by the shared symmetries between the microstructure geometry and the macroscopic stress state.  In this study, we focus on designing a square cell Fig.~\ref{fig:Illustration} (a)  {under  uniaxial stress} and  the corresponding IBZs  is illustrated in Fig.~\ref{fig:Illustration} (b).  Fig.~\ref{fig:Illustration} (c)  shows the corresponding buckling band diagrams calculated  as customary along the boundaries of IBZ and critical
buckling modes for the volume fraction of 20\%. It is seen that buckling strength is determined by a global shear mode with {$\sigma_c/E_1=0.00060$ with $\bar{E}/E_1=0.1074$}.  From the computational point of view, it is more convenient to perform LBA via $\left[ -\tau_h  \boldsymbol{K}_0   -    \boldsymbol{K}_{\sigma}  \right]\boldsymbol{\phi}_h = \boldsymbol{0} $, where $\tau_h =1/\lambda_h$.

\subsection{Design parameterization}
Based on the FEM discretization, an element-wise constant physical design variable, $\bar{\rho}_e$, is employed to represent the material distribution in element $e$, with  $\bar{\rho}_e=1$ and
$\bar{\rho}_e=0$ representing base material and void respectively.  As in~\cite{Thomsen2018}, to suppress the spurious buckling modes associated with the low stiffness elements, different interpolation schemes are employed for the elastic stiffness and stress stiffness using the solid isotropic material with penalization (SIMP) model \citep{Bendsoe1999}. The $\varepsilon$-relaxed approach from~\cite{Cheng1997}  is employed for the von Mises stress interpolation to avoid the stress singularity phenomenon. The interpolation  schemes are written as: 
\begin{equation} \label{Eq:2DCon}
	{E}_e = 
	\left\{
	\begin{array}{ll}
		\bar{\rho}^p_e(E_1-E_0)+E_0 & \text{for } \boldsymbol{K}_0,   \mathbf{f}_{\alpha} \\
		\bar{\rho}^p_e   {E}_1 & \text{for } \boldsymbol{K}_{\sigma} \ . \\
		\bar{\rho}_e/\left( \varepsilon (1-\bar{\rho}_e)+\bar{\rho}_e \right)   {E}_1   & \text{for }  {\sigma}^e_{vm}
	\end{array}
	\right.
\end{equation} 
Here $E_1$ is the base material Young's modulus,  $E_0=10^{-5} E_1$ represents void regions to avoid spurious modes located at the low stiffness region,   $p=3$ is chosen as the standard penalization factor. $\varepsilon=0.002$ is chosen for the von Mises interpolation. 

A hyperbolic tangent threshold projection is employed to generate physical design variables from the design variables,  ${\rho}_e$, to enhance  the discreteness  of the optimized design \citep{Wang2011}. This is given
\begin{equation} \label{Eq:2DProj}
	\bar{\rho}_e=\frac{\tanh{\left(\beta_1\eta\right)}+\tanh{\left(\beta_1\left(\widetilde{\rho}_e-\eta\right)\right)}}{\tanh{\left(\beta_1\eta\right)}+\tanh{\left(\beta_1\left(1-\eta\right)\right)}},
\end{equation}
where $\widetilde{\rho}_e$ is the filtered design variable calculated from the design variables, $\boldsymbol{\rho}$, using a PDE filter presented in~\cite{Lazarov2011b}.  When $\beta_1$ is big,   $\bar{\rho}_e\approx 1$ if $\widetilde{\rho}_e>\eta$  and $\bar{\rho}_e\approx0$ if $\widetilde{\rho}_e<\eta$. Hence the projection in Eq. \eqref{Eq:2DProj}  suppresses  gray element density regions induced by the PDE filter when  $\beta_1$ is big and ensures black-white designs when the optimization converges.  Moreover, it mimics the manufacturing process, and is used in the robust design formulation context~\citep{Wang2011}, where  manufacturing errors are taken into accounts by choosing different thresholds, $\eta$. In this study, the  maximum value of $\beta$ is chosen  to $\beta=8$  allowing a small amount of gray regions in {the} design to avoid  the mesh dependency of the maximum von Mises stress as discussed by~\cite{SilAagSig21}.

\subsection{Design problem formulation}
The KS function~\citep{Kreisselmeier1980} is employed  to aggregate the elemental von Mises stress ( $\sigma^e_{vm}$) to represent the maximum von Mises stress;   to aggregate  the considered eigenvalues for  given $\boldsymbol{k}$-vectors ($\tau_h\left(\boldsymbol{k}_l \right)$); to represent the material buckling strength; or to aggregate both quantities to represent both yielding and buckling strength, stated as
\begin{align}
	KS\left(\kappa_1  {\sigma^e_{vm}}/{\sigma_1}, \kappa_2  \tau_h\left(\boldsymbol{k}_l \right)   \right) &= \frac{1}{\zeta} \ln\left( \kappa_1  \sum_{e=1}^{N} e^{\zeta  \sigma^e_{vm}/\sigma_1}+  \kappa_2  \sum_{l=1}^{n_h} \sum_{h=1}^{m_l}e^{\zeta \left(  \tau_h\left( \boldsymbol{k}_l\right)\right) }\right), \qquad \kappa_1,\kappa_2 \in \left\lbrace 0;1 \right\rbrace. 
	\label{eq:KS_KSStress}
\end{align}
Here 
$\kappa_1=0$ and $\kappa_2=1$  aggregate the elemental von Mises stress,   $\kappa_1=1$ and $\kappa_2=0$ aggregate the eigenvalues  for  given $\boldsymbol{k}$-vectors,   $\kappa_1=1$ and $\kappa_2=1$ to aggregate both quantities.

The optimization problem for enhancing material stiffness and   strength can be formulated to minimize a weighted value of the KS function in Eq.~\eqref{eq:KS_KSStress} and $\bar{E}^{-1}$, stated as
\begin{align}
	\min\limits_{\boldsymbol{\rho}} \quad   &  \qquad  \left( \gamma_1	KS\left(\kappa_1 \sigma^e_{vm}/\sigma_1 ,\kappa_2  \tau_h\left(\boldsymbol{k}_l\right)   \right)    + \left( 1-\gamma_1 \right)\bar{E}^{-1}  \right) {E_1}, \qquad \kappa_1,\kappa_2 \in \left\lbrace 0;1 \right\rbrace      \nonumber  \\ 
	s.t. & \qquad  \left[ -\tau_h  \boldsymbol{K}_0 \left(\boldsymbol{k}_l \right) -    \boldsymbol{K}_{\sigma} \left( \boldsymbol{k}_l\right)  \right]\boldsymbol{\phi}_h = \boldsymbol{0} \nonumber \\
	& \qquad  \mathbf{K}_0 \boldsymbol{\chi}_{ \alpha} = \mathbf{f}_{\alpha}, \quad \alpha=1,2,3   \nonumber   \\ 
	& \qquad     {KS \left(\sigma^e_{vm}/{\sigma_1} \right)} {E_1} \leq  1/{\sigma}^* \ .  \\
	& \qquad  \bar{E}{/E_1}\geq E^*  \nonumber\\
	&  \qquad f={\sum_e v_e \bar{\rho}_e}/ {\sum_e v_e}   \leq f^* \nonumber \\
	&  \qquad  \boldsymbol{0} \leq \boldsymbol{\rho} \leq \boldsymbol{1} \nonumber
	\label{eq:opt}
\end{align}
Here   $\gamma_1$ is the weight, $\gamma_1=0$ and $\gamma_1=1$  represent stiffness and strength optimization, respectively. {$E_1$ in the objective  normalizes the microstructure strength and stiffness with respect to the base material Young's modulus.} $ {\sigma}^*$  is the {normalized} yield strength lower bound,   $ E^*$ is the {normalized} Young's modulus lower bound,   $v_e$ is the volume of element $e$, and  $f$ and $f^*$ are the actual and  the prescribed upper bound of the volume fraction in the microstructure.   

The sensitivity of a component, ${\bar{D}}_{ \alpha\beta}$, in the  effective elastic matrix  ${\bar{\boldsymbol{D}}}$    with respect to   $\bar{\rho}^e$, is written as
\begin{equation}\label{eq:CCSens}
	\frac{\partial {\bar{D}}_{\alpha\beta} } {\partial \bar{\rho}^e} = \frac{1}{|Y|}     \int_{Y^e}  \left(\tilde{\boldsymbol{\varepsilon}}_{\alpha} - {\hat{\mathbf{B}}}^e\boldsymbol{\chi}^e_{\alpha} \right)^T \frac{\partial  {\mathbf{D}}^e } {\partial \bar{\rho}^e}\left(\tilde{\boldsymbol{\varepsilon}}_{\beta} - {\hat{\mathbf{B}}}^e\boldsymbol{\chi}^e_{\beta} \right).  
\end{equation}
The sensitivities of  $\bar{E}$  can be analytical derived using Eqs. \eqref{eq:prop} and  \eqref{eq:CCSens}.

The sensitivities of elemental von Mises stress ($\sigma^e_{vm}$) with the respect to   $\bar{\rho}^e$  is calculated using the adjoint sensitivity analysis by  
\begin{equation}\label{eq:sigmaVM}
	\frac{\partial  \sigma^e_{vm}   } {\partial \bar{\rho}^e} =\frac{\left( \boldsymbol{\sigma}^e\right)^T\boldsymbol{M}}{\sqrt{\left( \boldsymbol{\sigma}^e\right)^T  \boldsymbol{M}\boldsymbol{\sigma}^e}}\left[   \frac{\partial {\mathbf{D}}^e } {\partial \bar{\rho}^e}   \Big( \mathbf{I} - {\hat{\mathbf{B}}}^e 
	\boldsymbol{X}^e
	\Big)   \boldsymbol{\varepsilon}_0+   {\mathbf{D}}^e     \Big( \mathbf{I} - {\hat{\mathbf{B}}}^e 
	\boldsymbol{X}^e
	\Big) \frac{\partial  \boldsymbol{\varepsilon}_0  } {\partial \bar{\rho}^e}\right]  + \sum_{\alpha=1}^3  \left( \boldsymbol{ \varphi}_ \alpha^e\right)^T               
	\left[ 
	\frac{\partial \boldsymbol{K}^e_{0}  }{\partial \bar{\rho}^e} \boldsymbol{\chi}^e_{ \alpha} - \frac{\partial \mathbf{f}^e_{\alpha}}{\partial \bar{\rho}^e} \right]. 
\end{equation}
Here  ${ \partial  \boldsymbol{\varepsilon}_0}/{\partial \bar{\rho}^e}$ can be directly derived using Eqs \eqref{eq:strain} and \eqref{eq:CCSens},   the adjoint vectors  $ \boldsymbol{\varPhi}=\left[\boldsymbol{\varphi_1},\boldsymbol{\varphi_2},\boldsymbol{\varphi}_3 \right] $ are obtained by solving 
\begin{equation}
	\boldsymbol{K}_{0}    \boldsymbol{\varPhi}=\sum_e \left[ \frac{\left( \boldsymbol{\sigma}^e\right)^T\boldsymbol{M}}{\sqrt{\left( \boldsymbol{\sigma}^e\right)^T  \boldsymbol{M}\boldsymbol{\sigma}^e}}  {\mathbf{D}}^e {\hat{\mathbf{B}}}^e\right] ^T \left[ \boldsymbol{\varepsilon}_0\right]^T.
\end{equation}

Assuming that the eigenvector is normalized, as $\left( \boldsymbol{\phi}_h \right)^H \boldsymbol{K}_0   \boldsymbol{\phi}_h = 1  $, the sensitivity of eigenvalue $\tau_h$ with respect to  $\bar{\rho}^e$ can be obtained via the adjoint sensitivity analysis as described below~\citep{Thomsen2018}, 
\begin{align}
	\frac{\partial \tau_h}{\partial \bar{\rho}^e}=&
	{\left( \boldsymbol{\phi}^e_h\right)}^H  \left[ -\tau_h \frac{\partial \boldsymbol{K}^e_{0}  }{\partial \bar{\rho}^e} -  \frac{\partial \boldsymbol{K}^e_{\sigma} } {\partial \bar{\rho}^e} \right]  \boldsymbol{\phi}^e_h + {\left( \boldsymbol{\phi}_h\right)}^H \left[  -  \frac{\partial \boldsymbol{K}_{\sigma} }{ \partial  \boldsymbol{\varepsilon}_0} \frac{ \partial  \boldsymbol{\varepsilon}_0}{\partial \bar{\rho}^e} \right] \boldsymbol{\phi}_h +\sum_{\alpha=1}^3\left( \boldsymbol{\psi}_ \alpha^e\right)^H \left[ 
	\frac{\partial \boldsymbol{K}^e_{0}  }{\partial \bar{\rho}^e} \boldsymbol{\chi}^e_{ \alpha} - \frac{\partial \mathbf{f}^e_{\alpha}}{\partial \bar{\rho}^e} \right]. 
	\label{eq:Sens}
\end{align} 
Here $\left( \right) ^H$ denotes the complex conjugate,  $\boldsymbol{\psi}_ \alpha$ is the adjoint vector corresponding to $\boldsymbol{\chi}_{\alpha}$, which is obtained by
\begin{align}
	\boldsymbol{K}_{0}    \boldsymbol{\psi}_\alpha=\sum_e  \left(\boldsymbol{\phi}^e_h\right)^H   \left[  \frac{\partial \boldsymbol{K}^e_{\sigma} }{ \partial  \boldsymbol{\chi}^e_\alpha}  \right] \boldsymbol{\phi}^e_h.
\end{align} 
The reader is referred to the work by~\cite{Thomsen2018} for the detailed calculation of $ \frac{\partial \boldsymbol{K}^e_{\sigma} }{ \partial  \boldsymbol{\chi}^e_ \alpha}$. 
{One of the advantages of using   aggregation functions is the uniqueness of the gradient
	of eigenvalues, even when eigenvalues are repeated as stated by~\cite{Torii2017}. }

The sensitivities of the objective and constraints with  respect to  a design variable, ${\rho}_e$,  are obtained using the chain rule. The optimization problem is implemented in  a flexible framework for large scale topology optimization~\citep{Aage2015} using the Portable Extensible Toolkit for Scientific computation (PETSc)~\citep{Balay2016} and Scalable Library for Eigenvalue Problem Computations  (SLEPc)~\citep{Hernandez2005}. The design is iteratively updated using the Method of Moving Asymptotes (MMA)~\citep{Svanberg1987} based on the gradients of the objective and constraints.  One case robust formulation in~\cite{Wang2021} is employed in this study, where the objective, yield and stiffness constrains is evaluated on an eroded microstructure generated using $0.5+ \Delta \eta$ with the  volume constraint working on the dilated microstructure generated with a threshold of $0.5- \Delta \eta$. The volume constraint is updated
every 20 iterations such that the volume constraint of the intermediate microstructure of $\eta = 0.5$ is satisfied. In this study, we choose $\Delta \eta=0.05$.

\section{Results}
The proposed optimization formulation is employed to design 2D square  microstructures with   45\textdegree-symmetry to achieve tunable stiffness and buckling strength response while considering three different volume fractions, i.e., $f^*=0.2,0.1, 0.05$. The unit cell is discretized by $512\times 512 $ $Q_{6}$ elements for $f^*=0.2$ and the resolution is doubled in order to enable the evolution of  structural hierarchy for $f^*=0.1$ and $f^*=0.05$. The filter radius is chosen to $r=0.03$ for  $f^*=0.2$ and it is reduced to $r=0.01$ for the 
two lower volume fractions. A square cell with 45\textdegree-symmetry is designed. {The initial microstructures (starting guesses) are chosen as the Young's modulus optimal microstructures with the prescribed volume fractions as shown in Fig.~\ref{fig:Illustration}}. 
The presented designs are the blueprints with $\eta=0.5$ unless otherwise stated. {Smooth and sharp boundaries are extracted as contour lines from the optimized grayscale designs  using the in-house code mentioned above, and subsequently evaluated in COMSOL using body-fitted meshes.}   The considered based material in the optimization  is  Pyrolytic Carbon (PC) with a Poisson's ratio of $\nu=1/3$ and  a relative yield strength of $\sigma_1/E_1=0.044$~\citep{Crook2020}.

\subsection{Validation of the proposed approach}

\begin{figure}[!htb]
	\includegraphics[width=1\textwidth]{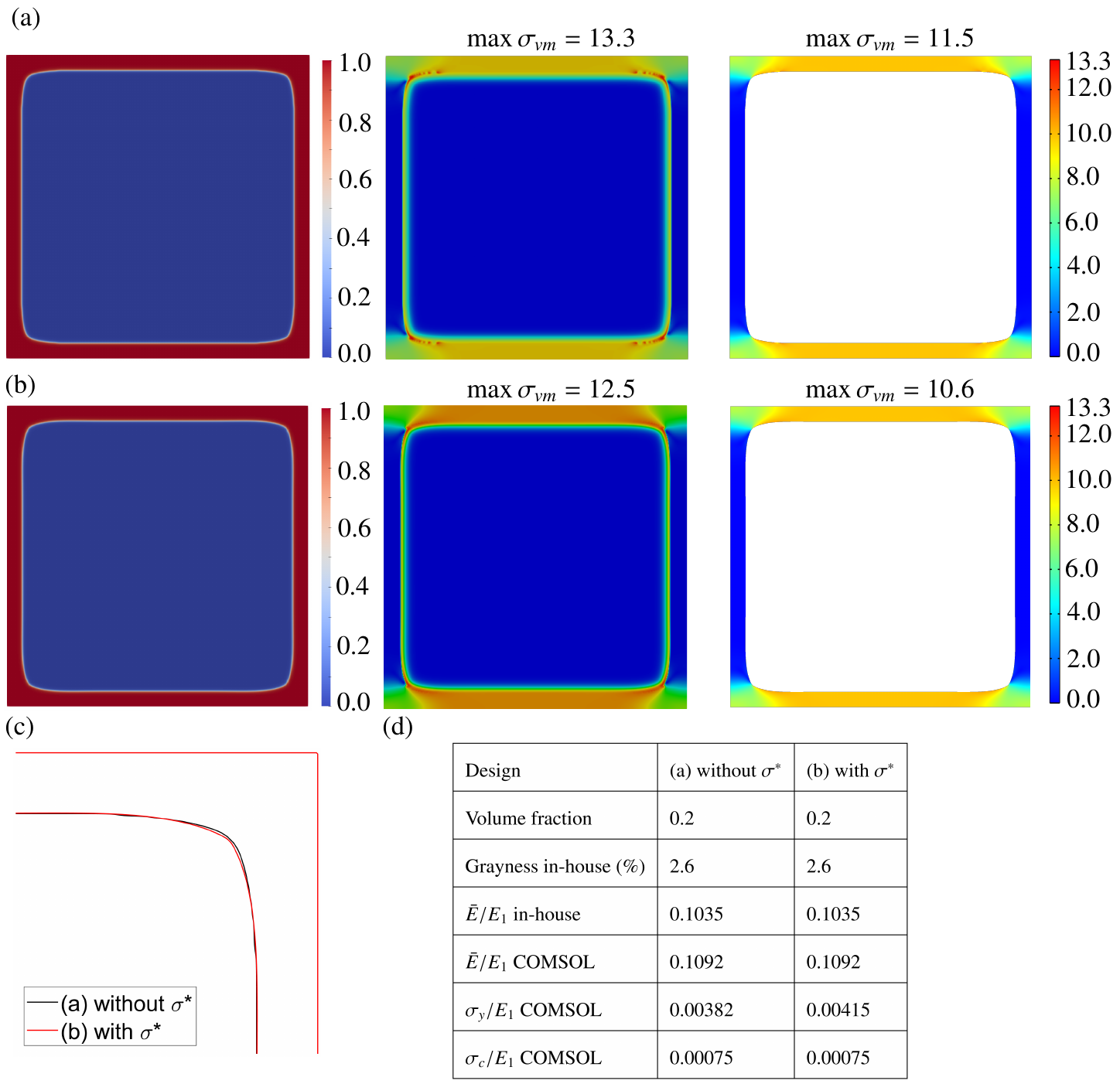} 
	{\caption{Optimized microstructures with maximized Young's modulus without (a) and with a yield strength constraint (b). Left: Optimized microstructures; {Middle: von Mises stress distribution evaluated at the element center using the in-house code with the optimized grayscale designs;  Right: von Mises stress distribution evaluated using the extracted designs in COMSOL with body-fitted meshes.} (c) Differences between the optimized designs. {(d) Performance summary of the two designs.} }	\label{fig:OptE}}
\end{figure}

In the first case, we  maximize microstructure Young's modulus without or with the yield strength constraint by setting  $\gamma_1 =0$. The corresponding normalized yield strength lower bound is set  $\sigma^*=1/400$.  The optimized  designs are extracted using the  0.5 isocontour and imported into COMSOL for post-evaluation using second-order body fitted meshes.   Fig. \ref{fig:OptE} (a) and (b) present the optimized microstructures obtained without and with the yield constraint, respectively. The left panels show the optimized microstructures, while the middle and right panels present the von Mises stress distribution evaluated using the in-house code and COMSOL. Both in-house code and COMSOL predict similar von Mises profiles and the same high von Mises stress regions. The intermediate physical densities in the optimized designs in the in-house code make the microstructure's perform slightly weaker  (i.e., lower effective Young's modulus) than the ones from COMSOL.  Hence the microstructures undergo bigger pre-strain  as indicated by Eq.~\eqref{eq:strain}  and exhibit larger maximum von Mises stress compared to the post-evaluation in COMSOL. These conclusions are verified by the von Mises stress distributions in both designs shown in Fig. \ref{fig:OptE}  (a) and (b).  

The  {rather insignificant} difference between the two designs is presented in  Fig. \ref{fig:OptE} (c) using the  0.5 isocontours of the 1/8 cell at the upper right corner. The optimized microstructures possess very similar shapes with small deviations at the inner corner. The inner corner in the optimized microstructure with the yield constraint exhibits slightly lower curvatures to fulfill the yield strength constraint. The properties of the optimized microstructures evaluated from the in-house code and COMSOL are summarized in Fig.  \ref{fig:OptE} (d). In the post-evaluation, both designs possess the same Young's modulus with 2\% to the theoretical Young's modulus upper bound, $E_u=f/(2-f)$. The material buckling evaluations of both designs show that global shear modes dominate buckling failure, and both designs possess higher buckling strength than the initial design in Fig. \ref{fig:Illustration}. The buckling strength enhancement is attributed to the round inner corner that makes the effect length of the member shorter than the initial design.   Comparison between yield and buckling strengths shows that optimized microstructures fail due to yield. 

The microstructure is subsequently  optimized by maximizing buckling strength with the same yield strength constraint, {which results in the same topology due to the limited design freedom provided by the yield strength bound}.  

In the rest of the paper, all the designs and corresponding performance will be presented using the well-defined post-evaluation in COMSOL,  unless otherwise stated.

\subsection{Strength-optimized microstructure}

\begin{figure}[!h]
	\includegraphics[width=1\textwidth]{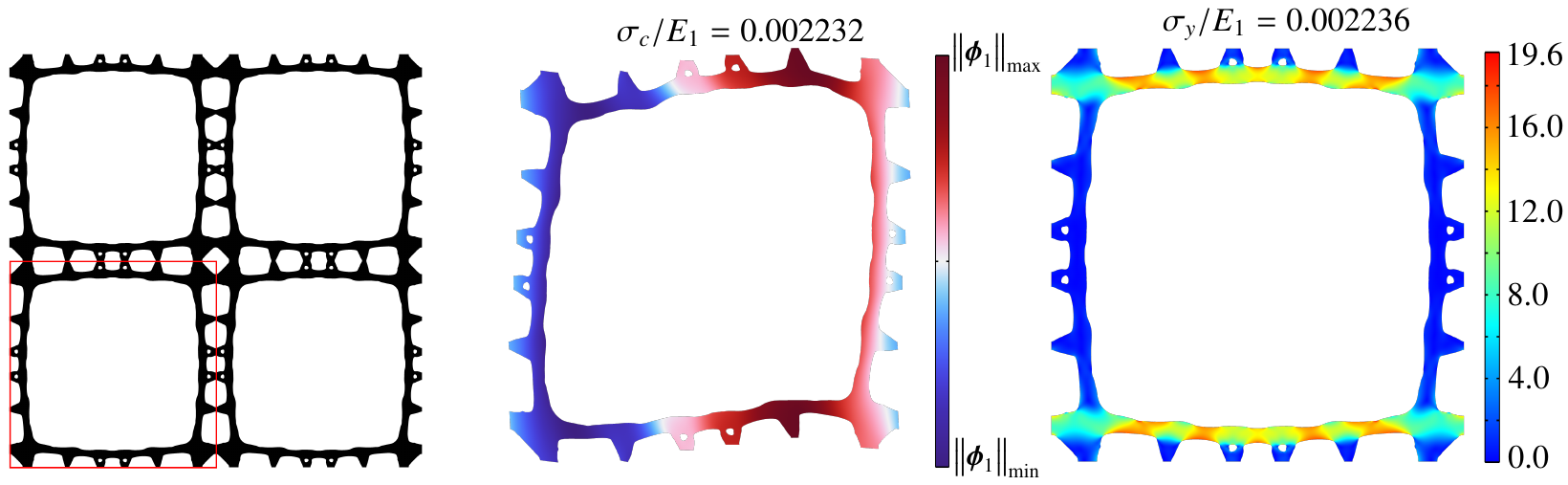} 
	{\caption{Optimized microstructure with maximized material strength.  Left: $2\times 2$ optimized cells, the red region highlights the optimized microstructure; Middle: Buckling failure mode with title representing buckling strength;  Right: von Mises stress distribution under uniaxial compression with title representing the yield strength.}	\label{fig:OptStr}}
\end{figure}

In the second case, a microstructure is optimized to maximize the material strength, i.e., maximizing the minimum value between the buckling and yield strengths by choosing $\kappa_1=1$,   $\kappa_2=1$    and  $\gamma_1 =1$.   Fig.~\ref{fig:OptStr} presents the optimized design and corresponding performance. Fig.~\ref{fig:OptStr}  (a) shows the optimized $2\times 2$ cells where the red box highlights the optimized microstructure.   Fig. ~\ref{fig:OptStr}  (b) and (c) present the corresponding critical buckling mode and von Mises stress distribution with titles showing the corresponding buckling and yield strengths. As in the previous designs, the critical buckling mode is a global shear mode. Compared to them, the optimized design develops  hierarchy to enhance the microstructure buckling strength. The hierarchy in the design enhances the effective width-to-length ratio of the members and leads to a higher buckling strength. On the other hand, it  reduces member stiffness and leads to a lower effective Young's modulus and yield strength. The optimized microstructure exhibits an effective Young's modulus of $\bar{E}/E_1= 0.08533$,  which is 21.9\% degradation compared to the stiffness-optimal design. For the considered base material, PC, the corresponding buckling and yield strengths manifest that the optimized microstructure fails simultaneously due to buckling and yield failures.

\subsection{Optimized microstructures with tunable stiffness/yield and buckling strength}

\begin{figure}[!h]
\includegraphics[width=1\textwidth]{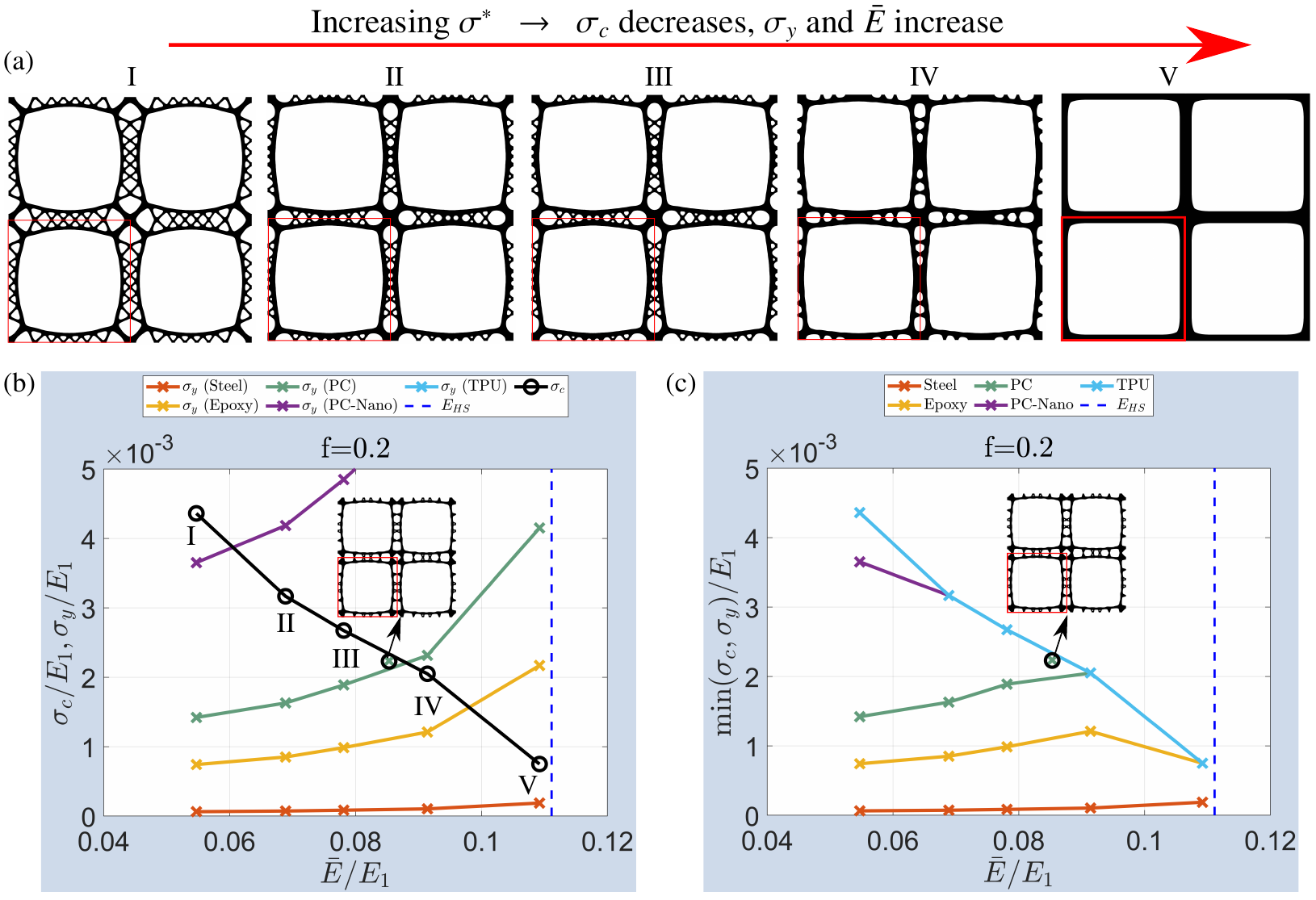}
	{\caption{Optimized microstructures for  {$f^*=0.2$}. (a) Optimized microstructures with increased yield strength bounds from left to right. (b) Yield/buckling strength versus Young's modulus performances for different constituent base materials.  Black line with circles:  Buckling strength vs. Young's modulus curve.   The other colored lines: Yield strength vs. Young's modulus.  Aqua: TPU; Purple: PC-Nano;  Teal: PC;  Yellow: Epoxy; Red: steel.  (c) Strength versus Young's modulus performances. }	\label{fig:OverAllV20}}
\end{figure}

We optimize  microstructure sets by maximizing  material buckling strength with  different yield strength bounds {for different volume fractions by choosing {$\kappa_1=0$,   $\kappa_2=1$}    and  $\gamma_1 =1$. }   The optimized design set {for $f^* =0.2$} is presented in Fig.~\ref{fig:OverAllV20} (a). As the required yield strength increases, the optimized microstructure becomes less hierarchical and converges to the {stiffness/yield-strength-optimal} one as shown in Fig.~\ref{fig:OptE} (b) for a large yield strength constraint. The optimized microstructures' performance is evaluated using different base materials ranging from low relative yield strength steel to high relative yield strength TPU . Tab.~\ref{Tab:base} summarizes the considered base materials  and {corresponding properties, including Young's modulus $E_1$, density $\rho_1$ and} relative yield strengths, i.e., the yield-strength-to-Young's-modulus ratio $\sigma_1/E_1$.

\begin{table}[!h]
 \caption{ {Considered constituent base materials and corresponding properties}~\citep{Crook2020,Andersen2021} }\label{Tab:base} 
 \centering
	{	\begin{tabular}{|c|c|c|c| }
			\hline
			&	$E_1$   	(GPa)  &  	$\rho_1$ ($\rm{kg/m^3}$) & $\sigma_1/E_1$ \\ 
			\hline
			Steel & 215   &7800 & 0.002  \\
			\hline
			Epoxy & 3.08 & 1400  & 0.023 \\
			\hline
			Pyrolytic Carbon (PC) & 62 & 1400 & 0.044 \\
			\hline
			Pyrolytic Carbon – nano  (PC-Nano) & 350    & 2600 &  0.113  \\ 
			\hline
			Thermoplastic Polyurethane (TPU)  &  0.012   & 1190   &  0.333 \\
			\hline
		\end{tabular}
	}
\end{table}

Fig.~\ref{fig:OverAllV20} (b) shows the normalized buckling strength ($\sigma_c$, black line with circles) and yield strengths for the considered base materials  ($\sigma_y$, colored lines with crosses, each color represents one base material) versus Young's modulus performance.  Fig.~\ref{fig:OverAllV20} (c) presents the optimized microstructures' strength versus Young's modulus performance.     For the considered base material in the optimization, i.e. PC, represented by the teal lines, it is seen that the increase of the yield strength bound leads to  optimized microstructures with increased yield strength and Young's modulus and reduced buckling strengths.  The relation between the effective Young's modulus and yield strength is observed in the microstructure set. There is a monotonic relation between the effective Young's modulus and yield strength for a given global stress situation.  It is explained by Eq.~\eqref{eq:strain} that the higher effective Young's modulus leads to a lower equivalent global strain for a given stress situation hence resulting in higher yield strength. For PC, the optimized design for material strength predicted by the optimized microstructure set is represented by the intersection between the buckling (the teal line with crosses) and yield strength (the black line with circles) curves. It is seen from Fig.~\ref{fig:OverAllV20} (b) that the strength-optimized design in Fig.~\ref{fig:OptStr}  { possesses a material strength very close to the  intersection point between the buckling curve (black) and the yield curve (teal), hence,  it provides the best combined yield and buckling  strength for PC. }  

\begin{figure}[!h]
 \includegraphics[width=1\textwidth]{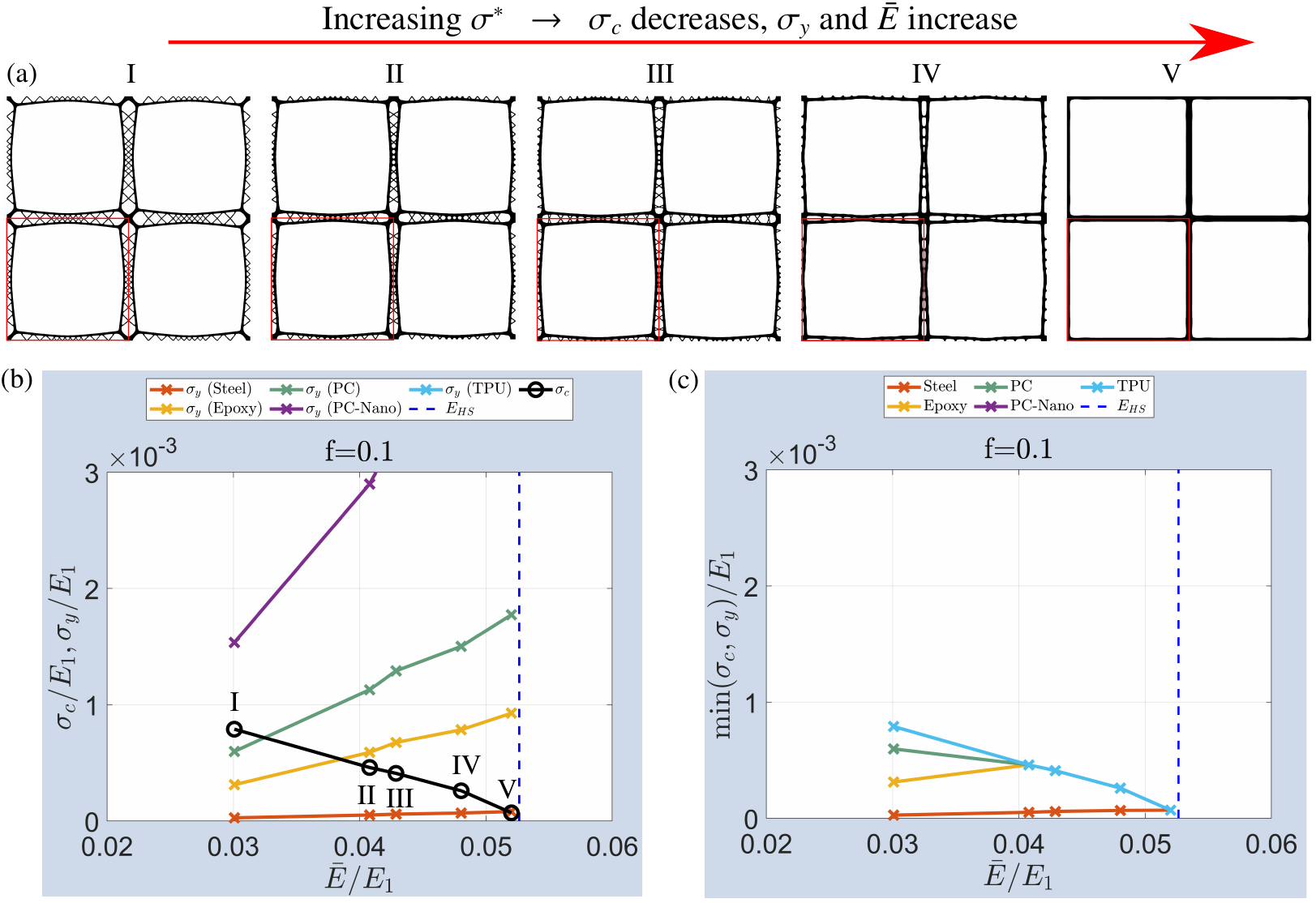} 
	{\caption{Optimized microstructures for  $f^*=0.1$. (a) Optimized microstructures with increased yield strength bound from left to right.   (b) Buckling/yield strength vs. Young's modulus performance considering different base materials. (c)  Microstructure strength vs. Young's modulus performances.}	\label{fig:OverAllV10}}
\end{figure}

The yield strength of the optimization set for TPU represented by the aqua-colored line is much bigger than the covered strength range in the plot; hence it does not appear in  Fig.~\ref{fig:OverAllV20} (b).  For TPU, all the optimized designs are dominated by buckling failure due to its high relative yield strength, and all the optimized designs are dominated by yield failure for steel due to its low relative yield strength. Failure mechanism transition between yield and buckling is observed for the rest of the considered base materials.  As the base material's relative yield strength increases, the transition between yield-dominated to buckling-dominated failure shifts to the left, i.e.,  to lower Young's modulus microstructures.  The strength-optimal microstructure switch to the optimized microstructure with  a lower Young's modulus as the relative yield strength of the base material increases as seen in Fig.~\ref{fig:OverAllV20} (c).

To further explore the optimized microstructure failure at the lower volume fractions, the microstructures are optimized for $f^*=0.1$ and $f^*=0.05$ by maximizing microstructure buckling strength for  different yield strength bounds.

Fig.~\ref{fig:OverAllV10} presents the optimized microstructure set for $f^*=0.1$  with increased yield strength bounds and the corresponding performances for different based materials. The geometrical evolutions in the design set show that the hierarchy vanishes as the required yield strength bound increases and the optimized microstructure configuration converges to the stiffness-optimal one for the largest yield strength bound. This observation is the same as in the previous case.   

Unlike the previous case, the optimized microstructure set fails due to buckling for  PC Nano and TPU. The failure mechanism transits from the yield-dominated to the buckling-dominated failure for the rest of the base materials when the optimized microstructure evolves from the buckling-optimal microstructure at the left to the stiffness/yield-optimal one at the right in Fig. (a). Compared to the design set for  $f^*=0.2$, The transition point shifts further to a lower Young's modulus range  with more hierarchy for $f^*=0.1$ for the same base material. 
\begin{figure}[!h]
 \includegraphics[width=1\textwidth]{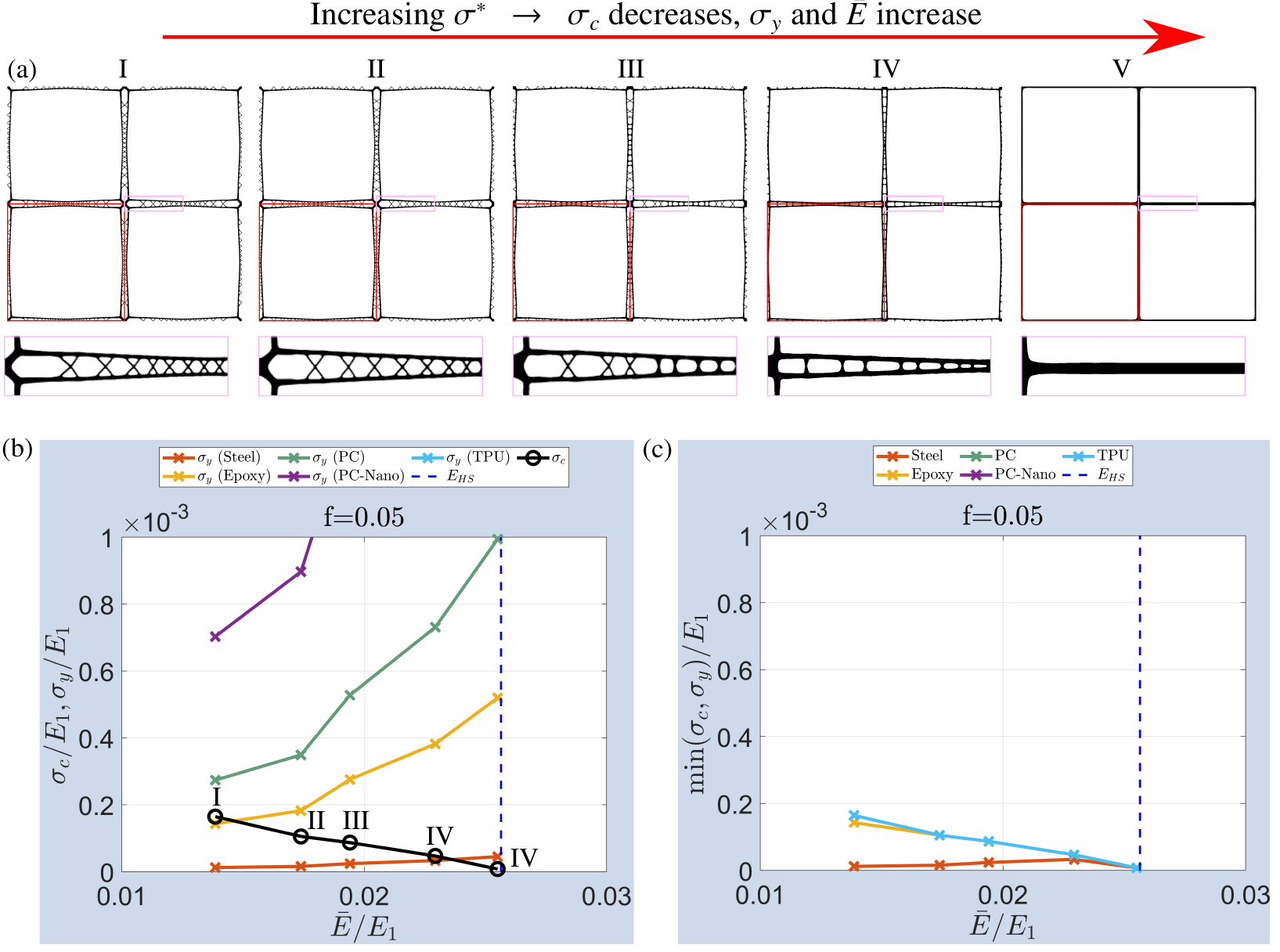} 
	{\caption{Optimized microstructures  for $f^*=0.05$. (a) Optimized microstructures with increased yield strength bound from left to right. Pink boxes  highlight the regions for enlarged views at the bottom.   (b) Buckling/yield strength vs. Young's modulus performances considering different base materials. (c)  Microstructure strength vs. Young's modulus performances.}	\label{fig:OverAllV5}}
\end{figure}

Fig.~\ref{fig:OverAllV5} summarizes the optimized microstructure set for  $f^*=0.05$ and corresponding performances with different base materials. A similar geometrical evolution is observed as in the previous two design sets. The optimized microstructure set fails due to buckling failure for PC, PC Nano, and TPU. Only the buckling-optimized design with the lowest yield strength bound fails due to yield strength for Epoxy, and the rest designs fail due to buckling. The failure mechanism transition is observed for steel. However, buckling failure dominates only the stiffness-optimal design, while yield failure dominates the rest designs. Compared with the previous two design sets for higher volume fractions, the transition between yield- and buckling-dominated failure shifts further towards the buckling-optimized designs.

The overall comparison between the three microstructure sets shows that the optimized microstructures are mainly dominated by buckling failure for the low volume fraction region, and they are dominated by yield failure for the high volume fraction region. It can be expected that all the optimized designs, from the buckling-optimal design to the stiffness/yield-optimal design, fail only due to buckling at an even lower volume fraction, and all the optimized designs have yield failure at much higher volume fractions. The low critical volume fraction is determined when the full hierarchical buckling optimal microstructure fails due to buckling. The high critical volume fraction is determined when the corresponding stiffness-optimal microstructure exhibits yield failure. The critical volume fractions are determined by the relative yield strength of the base material. For a  higher relative yield strength, the critical low volume fraction is higher  than the one with a lower one, while the high critical volume fraction is also higher. 

\begin{figure}[!h]
	 \includegraphics[width=1\textwidth]{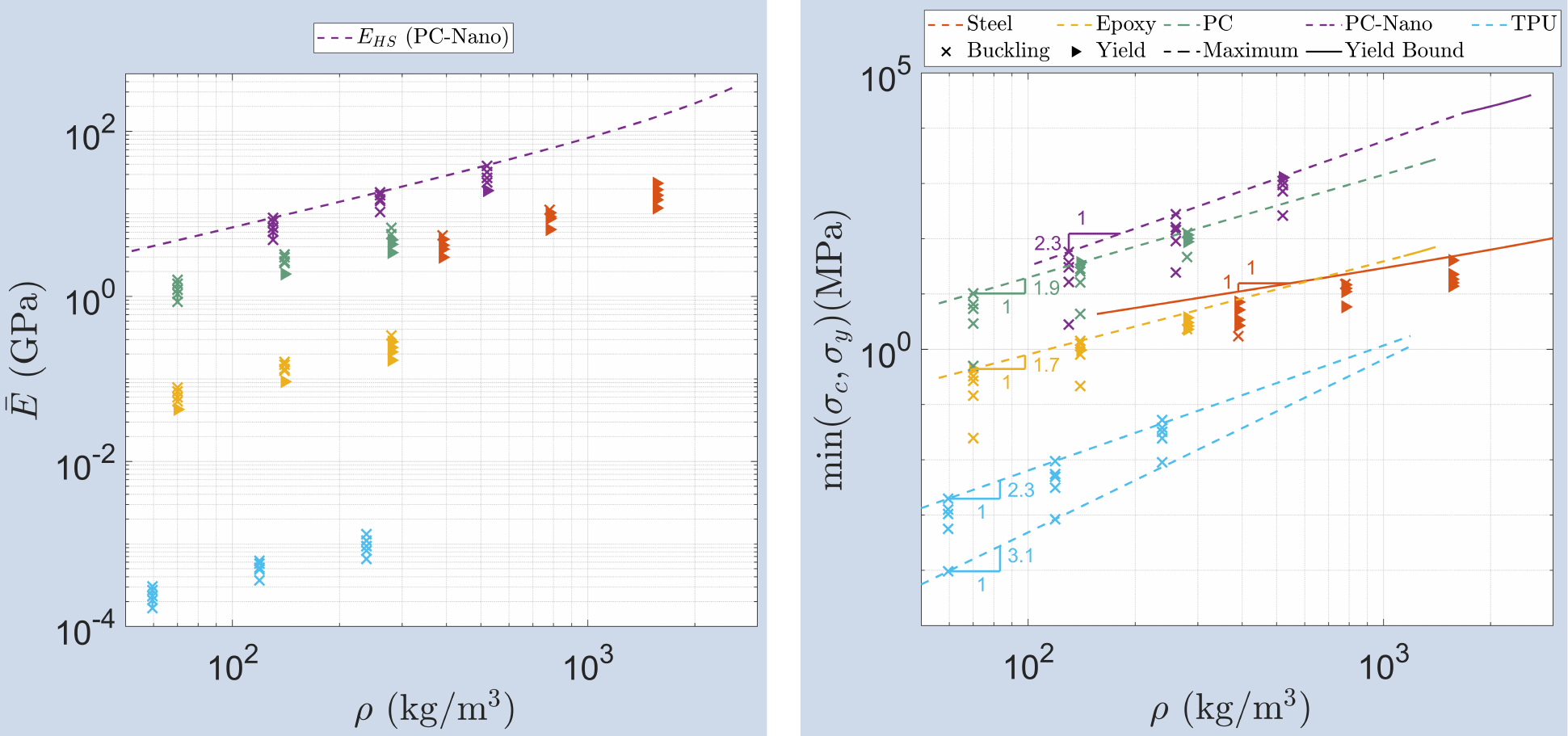} 
	{\caption{{Non-normalized properties of the optimized microstructures. Left: Young's modulus versus mass density with the dashed curve showing the Hashin–Shtrikman bound; Right: Strength versus mass density. The slopes indicate the polynomial order of the maximum strength lines to mass density.} }	\label{fig:OverAllTot}}
\end{figure}

{To provide additional insight into the role of the choice of different base materials, Fig.~\ref{fig:OverAllTot} summarizes the non-normalized material properties versus mass density for the optimized microstructures. PC-Nano possesses the highest Young's modulus and high relative yield strength (see Tab.~\ref{Tab:base}), hence, the optimized microstructures using PC-Nano exhibit the highest effective Young's modulus and strength while TPU results in the lowest effective Young's modulus and strength due to its low Young's modulus. As expected, the maximum Young's modulus curve of the optimized microstructures matches the Hashin–Shtrikman bound,  $E_{HS}=f/(2-f)E_1=\rho/(2\rho_1-\rho)E_1$; see the left panel. }

{The right panel presents the microstructure strength versus mass density, where $\times$'s indicate  microstructures failed due to buckling, while $\blacktriangle$'s indicate  microstructures failed due to yield. The dashed lines represent the maximum strength obtained from the optimized microstructures. By assuming a strength-mass-density relation of $ \min(\sigma_c,\sigma_y) = c_0 f^{n_0}=c_0 \left(\frac{\rho}{\rho_1} \right) ^{n_0}$, the order $n_0$ is obtained using the two points with lower densities and the orders for the different base materials are indicated in the strength plot as the line slopes. At low mass densities, the analytical yield strength upper bound follows the same relation as Young's modulus, i.e., $\sigma_y=f/(2-f)\sigma_1=\rho/(2\rho_1-\rho)\sigma_1$, represented by the solid lines in the plot and it possesses a linear relation with respect to mass density with $n_0=1$.  The minimum strength for TPU is controlled by  the simple Young's modulus-optimal microstructures with 1st order hierarchy (0th order indicates base material properties according to~\cite{Lakes1993}) and shows a close to cubic mass density dependency~\citep{Haghpanah2014}.   The maximum strength line for TPU  is controlled by the optimized  microstructures with  2nd order hierarchy, and  it has a polynomial order of $n_0=2.3$ that is slightly lower than the predicted one, 2.5, assuming self-hierarchy (see ~\cite{Andersen2021}), moreover with a higher coefficient $c_0$. As the relative yield strengths of the base materials increase, the mass density dependency order reduces and converges to $n_0=1$ for  steel that is governed by the yield upper bound. The maximum strength line  meets the yield strength upper bound at relatively high mass density for Epoxy, PC, and PC-Nano, while the one for TPU is solely governed by the maximum strength line, and the one for steel solely governed by the yield  bound. }

 \section{Conclusions}
 
 A general topology optimization formulation is proposed to tailor microstructure properties, including stiffness, yield, and buckling strength. The formulation is validated by designing square cell microstructures with maximal stiffness and/or strength. Numerical results for stiffness-optimal microstructures show that yield strength can be enhanced by a yield strength constraint without sacrificing the material stiffness and buckling strength. Compared to the stiffness-optimized microstructures, strength-optimized microstructures develop geometric hierarchy that enhances their buckling resistance. These strength-optimized microstructures have considerably higher strengths and fail simultaneously due to yield and buckling at a slight decrease in stiffness.

 Microstructure sets with volume fractions of 20\%, 10\%, and 5\% are systematically designed by maximizing buckling strength with different yield strength bounds. The geometric evolution of the optimized designs show that a lower yield strength bound allows microstructures to develop geometrical hierarchies, leading to higher buckling strength with decreased stiffnesses. Moreover, the effective Young's modulus and yield strength evolutions show a monotonic relation.    The optimized microstructures are post-evaluated for base materials with different relative yield strengths ranging from steel to TPU. The overall performances show that base materials with higher relative yield strength allow the strength-optimal designs to form geometrical hierarchies, leading to higher optimized strength with slightly decreased Young's modulus. Furthermore, as the design volume fraction decreases, the transition between yield- to buckling-dominated failures moves further toward the buckling-optimized designs. The overall performance indicates that buckling strength determines optimized strength at low-volume fractions and yield strength at high-volume fractions. The base material's relative yield strength determines the critical volume fractions for pure buckling-dominated or pure yield-dominated failures. For a volume fraction in between, the strength-optimized microstructure is the optimized one with simultaneous yield and buckling failures. Meanwhile, the failure mechanism switches from yield- to buckling-dominated when the optimized microstructure configurations evolve from the buckling-strength to the stiffness/yield-strength optimized one.  {The non-normalized properties of the optimized microstructures show that base materials with higher Young's modulus lead to both higher Young's modulus and strength. The polynomial order of the maximum strength lines of the optimized microstructures on the mass density reduces as the base material relative yield strength decreases,   reducing from 2.3 for TPU with buckling dominated to 1 for steel with yield dominated.}

\section*{Acknowledgments}
We acknowledge the financial support from the Villum Investigator Project InnoTop. 
%
%


\begin{thebibliography}{45}
	\expandafter\ifx\csname natexlab\endcsname\relax\def\natexlab#1{#1}\fi
	\providecommand{\url}[1]{\texttt{#1}}
	\providecommand{\href}[2]{#2}
	\providecommand{\path}[1]{#1}
	\providecommand{\DOIprefix}{doi:}
	\providecommand{\ArXivprefix}{arXiv:}
	\providecommand{\URLprefix}{URL: }
	\providecommand{\Pubmedprefix}{pmid:}
	\providecommand{\doi}[1]{\href{http://dx.doi.org/#1}{\path{#1}}}
	\providecommand{\Pubmed}[1]{\href{pmid:#1}{\path{#1}}}
	\providecommand{\bibinfo}[2]{#2}
	\ifx\xfnm\relax \def\xfnm[#1]{\unskip,\space#1}\fi
	\bibitem[{Meza et~al.(2015)Meza, Zelhofer, Clarke, Mateos, Kochmann, and
		Greer}]{Meza2015}
	\bibinfo{author}{L.~Meza}, \bibinfo{author}{A.~J. Zelhofer},
	\bibinfo{author}{N.~Clarke}, \bibinfo{author}{A.~J. Mateos},
	\bibinfo{author}{D.~Kochmann}, \bibinfo{author}{J.~Greer},
	\newblock \bibinfo{title}{Resilient 3d hierarchical architected metamaterials},
	\newblock \bibinfo{journal}{Proc Nat Acad Sci USA} \bibinfo{volume}{112}
	(\bibinfo{year}{2015}) \bibinfo{pages}{11502--11507}.
	\bibitem[{Zheng et~al.(2016)Zheng, Smith, Jackson, Moran, Cui, Chen, Ye, Fang,
		Rodriguez, and Weisgraber}]{Zheng2016}
	\bibinfo{author}{X.~Zheng}, \bibinfo{author}{W.~Smith},
	\bibinfo{author}{J.~Jackson}, \bibinfo{author}{B.~Moran},
	\bibinfo{author}{H.~Cui}, \bibinfo{author}{D.~Chen}, \bibinfo{author}{J.~Ye},
	\bibinfo{author}{N.~Fang}, \bibinfo{author}{N.~Rodriguez},
	\bibinfo{author}{T.~Weisgraber},
	\newblock \bibinfo{title}{Multiscale metallic metamaterials},
	\newblock \bibinfo{journal}{Nature materials} \bibinfo{volume}{15}
	(\bibinfo{year}{2016}) \bibinfo{pages}{1100}.
	\bibitem[{Sigmund(2000)}]{Sig00}
	\bibinfo{author}{O.~Sigmund},
	\newblock \bibinfo{title}{A new class of extremal composites},
	\newblock \bibinfo{journal}{Journal of the Mechanics and Physics of Solids}
	\bibinfo{volume}{48} (\bibinfo{year}{2000}) \bibinfo{pages}{397--428}.
	\bibitem[{Bends{\o}e and Sigmund(2003)}]{Bendsoe2003}
	\bibinfo{author}{M.~P. Bends{\o}e}, \bibinfo{author}{O.~Sigmund},
	\bibinfo{title}{{T}opology optimization: theory, methods and applications},
	\bibinfo{publisher}{Springer}, \bibinfo{address}{Berlin},
	\bibinfo{year}{2003}.
	\bibitem[{Osanov and Guest(2016)}]{OsaGue16}
	\bibinfo{author}{M.~Osanov}, \bibinfo{author}{J.~Guest},
	\newblock \bibinfo{title}{Topology optimization for architected materials
		design},
	\newblock \bibinfo{journal}{Annual Review of Materials Research}
	\bibinfo{volume}{46} (\bibinfo{year}{2016}) \bibinfo{pages}{211--233}.
	\bibitem[{Sigmund(1995)}]{Sigmund1995}
	\bibinfo{author}{O.~Sigmund},
	\newblock \bibinfo{title}{{T}ailoring materials with prescribed elastic
		properties},
	\newblock \bibinfo{journal}{Mechanics of Materials} \bibinfo{volume}{20}
	(\bibinfo{year}{1995}) \bibinfo{pages}{351--368}.
	\bibitem[{Guest and Prevost(2006)}]{Guest2006}
	\bibinfo{author}{J.~K. Guest}, \bibinfo{author}{J.~H. Prevost},
	\newblock \bibinfo{title}{{O}ptimizing multifunctional materials: {D}esign of
		microstructures for maximized stiffness and fluid permeability},
	\newblock \bibinfo{journal}{International Journal of Solids and Structures}
	\bibinfo{volume}{43} (\bibinfo{year}{2006}) \bibinfo{pages}{7028--7047}.
	\DOIprefix\doi{10.1016/j.ijsolstr.2006.03.001}.
	\bibitem[{Huang et~al.(2011)Huang, Radman, and Xie}]{Huang2011}
	\bibinfo{author}{X.~Huang}, \bibinfo{author}{A.~Radman}, \bibinfo{author}{Y.~M.
		Xie},
	\newblock \bibinfo{title}{{T}opological design of microstructures of cellular
		materials for maximum bulk or shear modulus},
	\newblock \bibinfo{journal}{Computational Materials Science}
	\bibinfo{volume}{50} (\bibinfo{year}{2011}) \bibinfo{pages}{1861--1870}.
	\DOIprefix\doi{10.1016/j.commatsci.2011.01.030}.
	\bibitem[{Andreassen et~al.(2014)Andreassen, Lazarov, and
		Sigmund}]{Andreassen2014}
	\bibinfo{author}{E.~Andreassen}, \bibinfo{author}{B.~Lazarov},
	\bibinfo{author}{O.~Sigmund},
	\newblock \bibinfo{title}{Design of manufacturable 3d extremal elastic
		microstructure},
	\newblock \bibinfo{journal}{Mechanics of Materials} \bibinfo{volume}{69}
	(\bibinfo{year}{2014}) \bibinfo{pages}{1--10}.
	\bibitem[{Berger et~al.(2017)Berger, Wadley, and McMeeking}]{Berger2017}
	\bibinfo{author}{J.~Berger}, \bibinfo{author}{H.~Wadley},
	\bibinfo{author}{R.~McMeeking},
	\newblock \bibinfo{title}{Mechanical metamaterials at the theoretical limit of
		isotropic elastic stiffness},
	\newblock \bibinfo{journal}{Nature} \bibinfo{volume}{543}
	(\bibinfo{year}{2017}) \bibinfo{pages}{533}.
	\bibitem[{Hashin(1962)}]{Hashin1962}
	\bibinfo{author}{Z.~Hashin},
	\newblock \bibinfo{title}{The elastic moduli of heterogeneous materials},
	\newblock \bibinfo{journal}{Journal of Applied Mechanics} \bibinfo{volume}{29}
	(\bibinfo{year}{1962}) \bibinfo{pages}{143--150}.
	\bibitem[{Ferrer et~al.(2021)Ferrer, Geoffroy-Donders, and
		Allaire}]{Ferrer2021}
	\bibinfo{author}{A.~Ferrer}, \bibinfo{author}{P.~Geoffroy-Donders},
	\bibinfo{author}{G.~Allaire},
	\newblock \bibinfo{title}{Stress minimization for lattice structures. part i:
		Micro-structure design},
	\newblock \bibinfo{journal}{Philosophical Transactions of the Royal Society A}
	\bibinfo{volume}{379} (\bibinfo{year}{2021}) \bibinfo{pages}{20200109}.
	\bibitem[{Collet et~al.(2018)Collet, Noël, Bruggi, and Duysinx}]{Collet2018}
	\bibinfo{author}{M.~Collet}, \bibinfo{author}{L.~Noël},
	\bibinfo{author}{M.~Bruggi}, \bibinfo{author}{P.~Duysinx},
	\newblock \bibinfo{title}{Topology optimization for microstructural design
		under stress constraints},
	\newblock \bibinfo{journal}{Structural and Multidisciplinary Optimization}
	\bibinfo{volume}{58} (\bibinfo{year}{2018}) \bibinfo{pages}{2677--2695}.
	\bibitem[{Coelho et~al.(2019)Coelho, Guedes, and Cardoso}]{Coelho2019}
	\bibinfo{author}{P.~G. Coelho}, \bibinfo{author}{J.~M. Guedes},
	\bibinfo{author}{J.~B. Cardoso},
	\newblock \bibinfo{title}{Topology optimization of cellular materials with
		periodic microstructure under stress constraints},
	\newblock \bibinfo{journal}{Structural and Multidisciplinary Optimization}
	\bibinfo{volume}{59} (\bibinfo{year}{2019}) \bibinfo{pages}{633--645}.
	\bibitem[{Alacoque et~al.(2021)Alacoque, Watkins, and Tamijani}]{Alacoque2021}
	\bibinfo{author}{L.~Alacoque}, \bibinfo{author}{R.~T. Watkins},
	\bibinfo{author}{A.~Y. Tamijani},
	\newblock \bibinfo{title}{Stress-based and robust topology optimization for
		thermoelastic multi-material periodic microstructures},
	\newblock \bibinfo{journal}{Computer Methods in Applied Mechanics and
		Engineering} \bibinfo{volume}{379} (\bibinfo{year}{2021})
	\bibinfo{pages}{113749}.
	\bibitem[{Cheng and Guo(1997)}]{Cheng1997}
	\bibinfo{author}{G.~D. Cheng}, \bibinfo{author}{X.~Guo},
	\newblock \bibinfo{title}{$\varepsilon$-relaxed approach in structural topology
		optimization},
	\newblock \bibinfo{journal}{Structural optimization} \bibinfo{volume}{13}
	(\bibinfo{year}{1997}) \bibinfo{pages}{258--266}. \URLprefix
	\url{https://doi.org/10.1007/BF01197454}.
	\bibitem[{Duysinx and Bendsøe(1998)}]{Duysinx1998}
	\bibinfo{author}{P.~Duysinx}, \bibinfo{author}{M.~P. Bendsøe},
	\newblock \bibinfo{title}{Topology optimization of continuum structures with
		local stress constraints},
	\newblock \bibinfo{journal}{International Journal for Numerical Methods in
		Engineering} \bibinfo{volume}{43} (\bibinfo{year}{1998})
	\bibinfo{pages}{1453--1478}.
	\bibitem[{Bruggi(2008)}]{Bruggi2008}
	\bibinfo{author}{M.~Bruggi},
	\newblock \bibinfo{title}{On an alternative approach to stress constraints
		relaxation in topology optimization},
	\newblock \bibinfo{journal}{Structural and Multidisciplinary Optimization}
	\bibinfo{volume}{36} (\bibinfo{year}{2008}) \bibinfo{pages}{125--141}.
	\URLprefix \url{https://doi.org/10.1007/s00158-007-0203-6}.
	\bibitem[{Le et~al.(2010)Le, Norato, Bruns, Ha, and Tortorelli}]{Le2010}
	\bibinfo{author}{C.~Le}, \bibinfo{author}{J.~Norato},
	\bibinfo{author}{T.~Bruns}, \bibinfo{author}{C.~Ha},
	\bibinfo{author}{D.~Tortorelli},
	\newblock \bibinfo{title}{Stress-based topology optimization for continua},
	\newblock \bibinfo{journal}{Structural and Multidisciplinary Optimization}
	\bibinfo{volume}{41} (\bibinfo{year}{2010}) \bibinfo{pages}{605--620}.
	\URLprefix \url{https://doi.org/10.1007/s00158-009-0440-y}.
	\bibitem[{Kreisselmeier and Steinhauser(1980)}]{Kreisselmeier1980}
	\bibinfo{author}{G.~Kreisselmeier}, \bibinfo{author}{R.~Steinhauser},
	\newblock \bibinfo{title}{Systematic control design by optimizing a vector
		performance index},
	\newblock \bibinfo{publisher}{Elsevier}, \bibinfo{year}{1980}, pp.
	\bibinfo{pages}{113--117}.
	\bibitem[{da~Silva et~al.(2021)da~Silva, Aage, Beck, and Sigmund}]{SilAagSig21}
	\bibinfo{author}{G.~da~Silva}, \bibinfo{author}{N.~Aage},
	\bibinfo{author}{A.~Beck}, \bibinfo{author}{O.~Sigmund},
	\newblock \bibinfo{title}{Local versus global stress constraint strategies in
		topology optimization: a comparative study},
	\newblock \bibinfo{journal}{International Journal for Numerical Methods in
		Engineering} \bibinfo{volume}{122} (\bibinfo{year}{2021})
	\bibinfo{pages}{6003--6036}. \DOIprefix\doi{10.1002/nme.6781}.
	\bibitem[{Neves et~al.(2002)Neves, Sigmund, and Bends{\o}e}]{Neves2002}
	\bibinfo{author}{M.~M. Neves}, \bibinfo{author}{O.~Sigmund},
	\bibinfo{author}{M.~P. Bends{\o}e},
	\newblock \bibinfo{title}{Topology optimization of periodic microstructures
		with a penalization of highly localized buckling modes},
	\newblock \bibinfo{journal}{International Journal for Numerical Methods in
		Engineering} \bibinfo{volume}{54} (\bibinfo{year}{2002})
	\bibinfo{pages}{809--834}.
	\bibitem[{Hassani and Hinton(1998)}]{Hassani1998a}
	\bibinfo{author}{B.~Hassani}, \bibinfo{author}{E.~Hinton},
	\newblock \bibinfo{title}{{A} review of homogenization and topology
		optimization {I} - homogenization theory for media with periodic structure},
	\newblock \bibinfo{journal}{Computers \& Structures} \bibinfo{volume}{69}
	(\bibinfo{year}{1998}) \bibinfo{pages}{707--717}.
	\bibitem[{Triantafyllidis and Schnaidt(1993)}]{Triantafyllidis1993}
	\bibinfo{author}{N.~Triantafyllidis}, \bibinfo{author}{W.~C. Schnaidt},
	\newblock \bibinfo{title}{Comparison of microscopic and macroscopic
		instabilities in a class of two-dimensional periodic composites},
	\newblock \bibinfo{journal}{Journal of the Mechanics and Physics of Solids}
	\bibinfo{volume}{41} (\bibinfo{year}{1993}) \bibinfo{pages}{1533--1565}.
	\bibitem[{Thomsen et~al.(2018)Thomsen, Wang, and Sigmund}]{Thomsen2018}
	\bibinfo{author}{C.~R. Thomsen}, \bibinfo{author}{F.~Wang},
	\bibinfo{author}{O.~Sigmund},
	\newblock \bibinfo{title}{Buckling strength topology optimization of 2d
		periodic materials based on linearized bifurcation analysis},
	\newblock \bibinfo{journal}{Computer Methods in Applied Mechanics and
		Engineering} \bibinfo{volume}{339} (\bibinfo{year}{2018})
	\bibinfo{pages}{115--136}.
	\bibitem[{Wang and Sigmund(2021)}]{Wang2021}
	\bibinfo{author}{F.~Wang}, \bibinfo{author}{O.~Sigmund},
	\newblock \bibinfo{title}{3d architected isotropic materials with tunable
		stiffness and buckling strength},
	\newblock \bibinfo{journal}{Journal of the Mechanics and Physics of Solids}
	\bibinfo{volume}{152} (\bibinfo{year}{2021}) \bibinfo{pages}{104415}.
	\bibitem[{Bluhm et~al.(2022)Bluhm, Christensen, Poulios, Sigmund, and
		Wang}]{Bluhm2021}
	\bibinfo{author}{G.~L. Bluhm}, \bibinfo{author}{K.~Christensen},
	\bibinfo{author}{K.~Poulios}, \bibinfo{author}{O.~Sigmund},
	\bibinfo{author}{F.~Wang},
	\newblock \bibinfo{title}{Experimental verication of a novel hierarchical
		lattice material with superiorbuckling strength},
	\newblock \bibinfo{journal}{APL Materials} \bibinfo{volume}{10}
	(\bibinfo{year}{2022}) \bibinfo{pages}{090701}.
	\bibitem[{Andersen et~al.(2021)Andersen, Wang, and Sigmund}]{Andersen2021}
	\bibinfo{author}{M.~N. Andersen}, \bibinfo{author}{F.~Wang},
	\bibinfo{author}{O.~Sigmund},
	\newblock \bibinfo{title}{On the competition for ultimately stiff and strong
		architected materials},
	\newblock \bibinfo{journal}{Materials \& Design} \bibinfo{volume}{198}
	(\bibinfo{year}{2021}) \bibinfo{pages}{109356}.
	\bibitem[{Cook et~al.(2002)Cook, Malkus, and Plesha}]{Cook2002}
	\bibinfo{author}{R.~Cook}, \bibinfo{author}{D.~Malkus},
	\bibinfo{author}{M.~Plesha}, \bibinfo{title}{{C}oncepts and {A}pplications of
		{F}inite {E}lement {A}nalysis}, \bibinfo{publisher}{Wiley},
	\bibinfo{address}{New York, N.Y.}, \bibinfo{year}{2002}.
	\bibitem[{Wilson et~al.(1973)Wilson, Taylor, Doherty, and
		Ghaboussi}]{Wilson1973}
	\bibinfo{author}{E.~L. Wilson}, \bibinfo{author}{R.~L. Taylor},
	\bibinfo{author}{W.~P. Doherty}, \bibinfo{author}{J.~Ghaboussi},
	\newblock \bibinfo{title}{Incompatible displacement models},
	\newblock \bibinfo{publisher}{Elsevier}, \bibinfo{year}{1973}, pp.
	\bibinfo{pages}{43--57}.
	\bibitem[{Wilson and Ibrahimbegovic(1990)}]{Wilson1990}
	\bibinfo{author}{E.~L. Wilson}, \bibinfo{author}{A.~Ibrahimbegovic},
	\newblock \bibinfo{title}{Use of incompatible displacement modes for the
		calculation of element stiffnesses or stresses},
	\newblock \bibinfo{journal}{Finite Elements in Analysis and Design}
	\bibinfo{volume}{7} (\bibinfo{year}{1990}) \bibinfo{pages}{229--241}.
	\bibitem[{Meza et~al.(2017)Meza, Phlipot, Portela, Maggi, Montemayor, Comella,
		Kochmann, and Greer}]{}
	\bibinfo{author}{L.~R. Meza}, \bibinfo{author}{G.~P. Phlipot},
	\bibinfo{author}{C.~M. Portela}, \bibinfo{author}{A.~Maggi},
	\bibinfo{author}{L.~C. Montemayor}, \bibinfo{author}{A.~Comella},
	\bibinfo{author}{D.~M. Kochmann}, \bibinfo{author}{J.~R. Greer},
	\newblock \bibinfo{title}{Reexamining the mechanical property space of
		three-dimensional lattice architectures},
	\newblock \bibinfo{journal}{Acta Materialia} \bibinfo{volume}{140}
	(\bibinfo{year}{2017}) \bibinfo{pages}{424--432}.
	\bibitem[{Brillouin(1953)}]{Brillouin1953}
	\bibinfo{author}{L.~Brillouin}, \bibinfo{title}{Wave Propagation in Periodic
		Structures: Electric Filters and Crystal Lattices},
	\bibinfo{publisher}{Dover, New York}, \bibinfo{year}{1953}.
	\bibitem[{Geymonat et~al.(1993)Geymonat, Müller, and
		Triantafyllidis}]{Geymonat1993}
	\bibinfo{author}{G.~Geymonat}, \bibinfo{author}{S.~Müller},
	\bibinfo{author}{N.~Triantafyllidis},
	\newblock \bibinfo{title}{Homogenization of nonlinearly elastic materials,
		microscopic bifurcation and macroscopic loss of rank-one convexity},
	\newblock \bibinfo{journal}{Archive for rational mechanics and analysis}
	\bibinfo{volume}{122} (\bibinfo{year}{1993}) \bibinfo{pages}{231--290}.
	\bibitem[{Bends{\o}e and Sigmund(1999)}]{Bendsoe1999}
	\bibinfo{author}{M.~P. Bends{\o}e}, \bibinfo{author}{O.~Sigmund},
	\newblock \bibinfo{title}{{M}aterial interpolation schemes in topology
		optimization},
	\newblock \bibinfo{journal}{Archive of Applied Mechanics} \bibinfo{volume}{69}
	(\bibinfo{year}{1999}) \bibinfo{pages}{635--654}.
	\bibitem[{Wang et~al.(2011)Wang, Lazarov, and Sigmund}]{Wang2011}
	\bibinfo{author}{F.~Wang}, \bibinfo{author}{B.~S. Lazarov},
	\bibinfo{author}{O.~Sigmund},
	\newblock \bibinfo{title}{On projection methods, convergence and robust
		formulations in topology optimization},
	\newblock \bibinfo{journal}{Structural and Multidisciplinary Optimization}
	\bibinfo{volume}{43} (\bibinfo{year}{2011}) \bibinfo{pages}{767--784}.
	\bibitem[{Lazarov and Sigmund(2011)}]{Lazarov2011b}
	\bibinfo{author}{B.~S. Lazarov}, \bibinfo{author}{O.~Sigmund},
	\newblock \bibinfo{title}{{F}ilters in topology optimization based on
		{H}elmholtz‐type differential equations},
	\newblock \bibinfo{journal}{International Journal for Numerical Methods in
		Engineering} \bibinfo{volume}{86} (\bibinfo{year}{2011})
	\bibinfo{pages}{765--781}.
	\bibitem[{Torii and Faria(2017)}]{Torii2017}
	\bibinfo{author}{A.~J. Torii}, \bibinfo{author}{J.~R.~d. Faria},
	\newblock \bibinfo{title}{Structural optimization considering smallest
		magnitude eigenvalues: a smooth approximation},
	\newblock \bibinfo{journal}{Journal of the Brazilian Society of Mechanical
		Sciences and Engineering} \bibinfo{volume}{39} (\bibinfo{year}{2017})
	\bibinfo{pages}{1745--1754}.
	\bibitem[{Aage et~al.(2015)Aage, Andreassen, and Lazarov}]{Aage2015}
	\bibinfo{author}{N.~Aage}, \bibinfo{author}{E.~Andreassen},
	\bibinfo{author}{B.~S. Lazarov},
	\newblock \bibinfo{title}{Topology optimization using petsc: An easy-to-use,
		fully parallel, open source topology optimization framework},
	\newblock \bibinfo{journal}{Structural and Multidisciplinary Optimization}
	\bibinfo{volume}{51} (\bibinfo{year}{2015}) \bibinfo{pages}{565--572}.
	\bibitem[{Balay et~al.(2016)Balay, Abhyankar, Adams, Brown, Brune, Buschelman,
		Dalcin, Eijkhout, Gropp, Kaushik, Knepley, McInnes, Rupp, Smith, Zampini,
		Zhang, and Zhang}]{Balay2016}
	\bibinfo{author}{S.~Balay}, \bibinfo{author}{S.~Abhyankar},
	\bibinfo{author}{M.~F. Adams}, \bibinfo{author}{J.~Brown},
	\bibinfo{author}{P.~Brune}, \bibinfo{author}{K.~Buschelman},
	\bibinfo{author}{L.~Dalcin}, \bibinfo{author}{V.~Eijkhout},
	\bibinfo{author}{W.~D. Gropp}, \bibinfo{author}{D.~Kaushik},
	\bibinfo{author}{M.~G. Knepley}, \bibinfo{author}{L.~C. McInnes},
	\bibinfo{author}{K.~Rupp}, \bibinfo{author}{B.~F. Smith},
	\bibinfo{author}{S.~Zampini}, \bibinfo{author}{H.~Zhang},
	\bibinfo{author}{H.~Zhang}, \bibinfo{title}{PETSc Users Manual},
	\bibinfo{type}{Technical Report} \bibinfo{number}{ANL-95/11 - Revision 3.7},
	\bibinfo{year}{2016}.
	\bibitem[{Hernandez et~al.(2005)Hernandez, Roman, and Vidal}]{Hernandez2005}
	\bibinfo{author}{V.~Hernandez}, \bibinfo{author}{J.~E. Roman},
	\bibinfo{author}{V.~Vidal},
	\newblock \bibinfo{title}{Slepc: A scalable and flexible toolkit for the
		solution of eigenvalue problems},
	\newblock \bibinfo{journal}{ACM Transactions on Mathematical Software (TOMS)}
	\bibinfo{volume}{31} (\bibinfo{year}{2005}) \bibinfo{pages}{351--362}.
	\bibitem[{Svanberg(1987)}]{Svanberg1987}
	\bibinfo{author}{K.~Svanberg},
	\newblock \bibinfo{title}{The method of moving asymptotes--a new method for
		structural optimization},
	\newblock \bibinfo{journal}{International journal for numerical methods in
		engineering} \bibinfo{volume}{24} (\bibinfo{year}{1987})
	\bibinfo{pages}{359--373}.
	\bibitem[{Crook et~al.(2020)Crook, Bauer, Guell~Izard, Santos~de Oliveira,
		Martins de Souza~e Silva, Berger, and Valdevit}]{Crook2020}
	\bibinfo{author}{C.~Crook}, \bibinfo{author}{J.~Bauer},
	\bibinfo{author}{A.~Guell~Izard}, \bibinfo{author}{C.~Santos~de Oliveira},
	\bibinfo{author}{J.~Martins de Souza~e Silva}, \bibinfo{author}{J.~B.
		Berger}, \bibinfo{author}{L.~Valdevit},
	\newblock \bibinfo{title}{Plate-nanolattices at the theoretical limit of
		stiffness and strength},
	\newblock \bibinfo{journal}{Nature communications} \bibinfo{volume}{11}
	(\bibinfo{year}{2020}) \bibinfo{pages}{1--11}.
	\bibitem[{Lakes(1993)}]{Lakes1993}
	\bibinfo{author}{R.~Lakes},
	\newblock \bibinfo{title}{Materials with structural hierarchy},
	\newblock \bibinfo{journal}{Nature} \bibinfo{volume}{361}
	(\bibinfo{year}{1993}) \bibinfo{pages}{511--515}.
	\bibitem[{Haghpanah et~al.(2014)Haghpanah, Papadopoulos, Mousanezhad,
		Nayeb-Hashemi, and Vaziri}]{Haghpanah2014}
	\bibinfo{author}{B.~Haghpanah}, \bibinfo{author}{J.~Papadopoulos},
	\bibinfo{author}{D.~Mousanezhad}, \bibinfo{author}{H.~Nayeb-Hashemi},
	\bibinfo{author}{A.~Vaziri},
	\newblock \bibinfo{title}{Buckling of regular, chiral and hierarchical
		honeycombs under a general macroscopic stress state},
	\newblock \bibinfo{journal}{Proc. R. Soc. a} \bibinfo{volume}{470}
	(\bibinfo{year}{2014}) \bibinfo{pages}{20130856}.
	
\end{thebibliography}
\end{document}